# Simultaneous Single Crystal Growth and Segregation of Ni-Rich Cathode Enabled by Nanoscale Phase Separation for Advanced Lithium-Ion Batteries

Yujing Bi[1], Yaobin Xu[1], Ran Yi[1], Dianying Liu[1], Peng Zuo[1], Jiangtao Hu[1], Qiuyan Li[1], Jing Wu[1], Chongmin Wang[1], Sha Tan[2], Enyuan Hu[2], Jingnan Li[3], Rebecca O'Toole[3], Liu Luo[3], Xiaoguang Hao[3], Subramanian Venkatachalam[3], Job Rijssenbeek[3], Jie Xiao[1,4]*

[1] Energy & Environmental Directorate, Pacific Northwest National Laboratory, Richland, WA 99352, USA

[2] Chemistry Division, Brookhaven National Laboratory, Upton, NY 11973, USA

[3] Albemarle, Kings Mountain, NC 28086, USA

[4] Department of Materials Science & Engineering, University of Washington, Seattle, WA98101, USA

*Correspondence to: jie.xiao@pnnl.gov

**Abstract:** Synthesis of high-performance single crystal $LiNi_{0.8}Mn_{0.1}Co_{0.1}O_2$ (NMC811) in the absence of molten salt is challenging with no success yet. An innovative drop-in approach is discovered to synthesize single crystal NMC811 by controlling the morphology of transition metal hydroxide $TM(OH)_2$ precursors followed by a simple decomposition step to form transition metal oxide (TMO) intermediates. Ni redistribution in TMO, as a result of the concurrent formation of mixed spinel and rock salt phases, helps deagglomerate the later formed NMC811 clusters of single crystals. As-prepared single crystal NMC811 is validated in a 2Ah pouch cell demonstrating 1000 stable cycling. The fundamentally new reaction mechanism of single crystal growth and segregation without molten salt provides a new direction towards cost-efficient manufacturing of single crystal NMC811 cathode for advanced lithium-based batteries.

Single crystal Ni-rich cathodes such as LiNi$_{0.8}$Mn$_{0.1}$Co$_{0.1}$O$_2$ (NMC811) are of great importance because of their potentials to overcome the materials challenges of high energy Ni-rich cathode materials for next generation lithium-based batteries for electrical vehicles (EVs) [1,2]. However, Ni-rich NMC cathode has its own problems such as moisture sensitivity [3,4], cracking [5,6] and gas generation [7,8] that plague its large-scale application in EV batteries. The challenges of NMC811 polycrystals mainly initiate from grain boundaries that can potentially be eliminated or mitigated in single crystal format [9-11]. While the definition of single crystal for Ni rich NMC is still arguable, in general, micron sized NMC particles without significant agglomeration can be considered as single crystals. Slight agglomeration of crystals sometimes is unavoidable. But the primary particle size is at micron level and do not aggregate together thus eliminating the negative effects from grain boundaries.

Synthesis of high-performance single crystal Ni-rich NMC is challenging. As more Ni content is present in NMC, a lower calcination temperature is preferred because of Ni reduction at elevated temperatures, while high temperatures favor single crystal growth [12]. In addition, the long-time sintering process necessary to form single crystals facilitates lithium salt evaporation at high temperatures leading to non-stoichiometry in NMC811 with poor performances. Therefore, molten salt is usually employed as the reaction media to promote growth of single crystal NMC811 [2,13,14]. To manufacture single crystal NMC811 materials on a large scale, solid state synthesis without using any molten salt is of great importance from cost aspect which has not been a success yet [15].

This work explores an innovative drop-in approach for synthesis of high-performance single crystal NMC811 that is readily adaptable for industry manufacturing. The reaction mechanism of single crystal growth is investigated to understand the synthesis-morphology-performance relationship in growing single crystal NMC811 in the absence of molten salts. The scaled single crystal NMC811 is then further validated in a 2Ah Li-ion pouch cell which demonstrates 1000 stable cycling, confirming the feasibility of this new synthesis approach. This work provides new insights broadly applicable for cost-efficient large-scale synthesis of single crystals for different advanced battery technologies to accelerate deep decarbonization process.

**Results and Discussion**
Commercial polycrystalline NMC is usually synthesized by using transition metal hydroxide TM(OH)$_2$ as the precursor (Fig.1 and Fig.s1a) which is firstly prepared through coprecipitation reaction [16]. Then lithium salt such as LiOH is mixed with TM(OH)$_2$ followed by high temperature calcination to form layered NMC. The as-prepared polycrystalline NMC usually consists of nano-sized primary particles which agglomerate into micron sized secondary particles (Fig.s1b), i.e., polycrystalline NMC. If single crystal is desired, molten salt media is usually needed to promote crystal growth [2] without sacrificing final performances. Increasing the sintering temperature helps forming micron sized crystals which, however, significantly agglomerate together. A post-synthesis grinding process is needed to break down those agglomerated crystals which demonstrates inferior performances than their polycrystal counterparts[17]. This group recently found that simply converting TM(OH)$_2$ hydroxide (Fig.s1a) to transition metal oxide (TMO) intermediate precursor (Fig.1 & Fig.s1c) enables direct synthesis of single crystal NMC811 (Fig.1 and Fig.s1d). The removal of molten salt in synthesis provides significant convenience and reduces manufacturing cost for industry single crystal processing. There are two critical parameters that need careful control when applying this new approach: (1) decomposition of TM(OH)$_2$ into TMO to promote crystal growth and segregation in NMC811, and (2) tailoring microstructures of TM(OH)$_2$ to further minimize the agglomeration of later formed NMC811 single crystals.

The necessity of converting TM(OH)$_2$ into TMO is based on the observation that NMC811 usually copies its parent precursor structure. For example, the state-of-art synthesis of polycrystal NMC811 takes advantage of agglomerated TM(OH)$_2$ (Fig.s1a) precursor and displays a secondary structure (Fig.s1b) consisting of assembled nano-sized primary NMC particles. If micron sized single crystal NMC811 is desired, the primary particles size in the precursors need to be increased first. However, conventional solution-based precipitation reaction only creates aggregated nanoplates of TM(OH)$_2$. A simple approach to increase the primary particle size in the precursors is to decompose TM(OH)$_2$ at high temperatures to form oxide intermediates in which submicron sized TMO primary particles readily form (Fig.s1c).

A closer look at TM(OH)$_2$ precipitates reveals that numerous ultra-thin "nano-plates" agglomerate into spherical particles (Fig.s1a) in which pores and voids are observed across the entire particles (Fig.2a). After high temperature decomposition, TM(OH)$_2$ nano plates evolve into submicron sized TMO crystals which still resemble secondary structures (Fig.s1c and Fig.2b). TM(OH)$_2$ precursor is $\beta$-Ni(OH)$_2$ phase (P$\bar{3}$m1 pace group, Fig.s2a) and decomposes into TMO which is a mixed phase of rock salt (Fm$\bar{3}$m) and spinel (Fd$\bar{3}$m) structures (Fig.2c). Rock salt structure is the major phase (see HADDF image and BF in Fig.s3) in TMO, consistent with XRD observations (Fig.s2b). After reaction with LiOH, both rock salt and spinel phases of TMO are converted to the same layered NMC811(Fig.2f&2g, Fig.s2a) with an average crystal size of ca.1 µm (Fig.s1d). Cross-sectional image of single crystal indicates dense structures without any grain boundary (Fig.2d). The surface of as-prepared single crystal NMC811 is also quite clean with a few nanometers of cation mixing region observed on crystal surface (Fig.s4). Transition metal elements Ni, Mn and Co are uniformly distributed across entire crystals with desired stoichiometry (Fig.2e).

During the heating process of TM(OH)$_2$, from 500°C to 600°C, there is no noticeable morphological change of as-derived TMO (Fig.3a1-3b1). But TEM image of 600°C-dervied TMO (Fig.3b2) uncovers a mixture of nanoplates and spherical nanoparticles, while 500°C TMO is dominated by nanoplates (Fig.3a1-a2) similar to its TM(OH)$_2$ precursor (Fig.s1a). Rock salt and spinel phases of TMO show up at as low as 500°C (Fig.3a3 and XRD in Fig.s2b) and are always accompanying each other during the entire decomposition process (Fig.3b3-3e3). As decomposition temperature increases to 700°C, more spherical primary particles form (Fig.3c1-3c2) and continue to grow to submicron sizes at 800°C (Fig.3d1-d2) and 900°C (Fig.3e1-e2), accompanied by the enhanced crystallinity of TMO (Fig.3c3, 3d3, 3e3 and Fig.s5-s10). Further increasing the decomposition temperature to 1000°C forms bigger primary TMO particles of ca. 2µm (Fig.s11). However, oxygen loss (Fig.s12) also happens aggressively at such a high temperature leading to significant oxygen deficiency in the lattice structures of 1000°C-derived TMO. These oxygen deficiencies formed in 1000°C-decomposed TMO are difficult to be fully restored later during crystal growth reaction or annealing process leading to the observed inferior electrochemical performances of NMC811 single crystals (Fig.s13). Ni, Mn and Co are homogeneously distributed in 500°C-derived TMO (Fig.3a4). As decomposition temperature increases, the distribution of Ni (colored by green in EDS images of Fig.3) has been found to keep evolving and becomes enriched near TMO surfaces (Fig.3b4-3c4). Mn (colored by blue in EDS images of Fig.3) and Co (colored by pink in EDS images of Fig.3) are always accompanying each other without segregation (Fig.3a4-3e4) regardless of decomposition temperatures (Fig.s5-s9). This is related to the stoichiometry difference and uneven distribution of rock salt and spinel phases in TMO that drives Ni re-distribution. Rock salt phase has more Ni than in the spinel phase in TMO, while Mn and Co ratio and distribution are consistent in both phases. At 800°C, the segregation of Ni becomes more prominent (Fig.3d4) on some of the TMO primary particles, leading to the capture of rock salt phase region (green area in Fig.3d4) covering the surface of spinel phase (pink/bule region in Fig.3d4). This phenomenon is further amplified by 900°C-developed TMO which displays Ni-rich rock salt phase on surfaces and phase boundaries of many TMO particles (the scale bar in Fig.3e4 is increased to include more TMO particles for comparison). After reacting with LiOH, however, both phases are converted to layered $\alpha$-NaFeO$_2$ phase (Fig.s2a). A trace amount of Ni relocation along TMO grain boundaries (Fig.3e4) loosens the agglomerated structures which facilitates the formation of separated individual crystals after reaction with Li salt, leaving behind a nano-scale region of Li$^+$/Ni$^{2+}$ cation mixing region on NMC811 crystal surface (Fig.s4).

In addition to the TM(OH)$_2$-to-TMO conversion step, the original particle size of TM(OH)$_2$ precipitate also plays a significant role in the morphology and properties of as-prepared NMC811 single crystals. TM(OH)$_2$ with different secondary particle sizes of 3.6, 4.6, 6.7 µm (see particle size distribution in Fig.s14) are prepared for comparison. Note that the primary particles in all three TM(OH)$_2$ are similar residing in the nano range and the only difference among them is the dimension of their agglomerated secondary particles (Figs.4a1, 4b1 and 4c1). After decomposition, the TMO intermediates still inherit the agglomerated structures (Fig.4a2, 4b2 and 4c2) of their corresponding TM(OH)$_2$ precursors. However, the primary particles in each TMO have grown into submicron-sized crystals, while in TM(OH)$_2$ the primary particles are nanoplates. The three TMO intermediates further react with LiOH at 800°C to form NMC811, which now demonstrate different morphological structures with TMO or TM(OH)$_2$. For NMC811 single crystals derived from 3.6 (Fig.4a3) or 4.6 µm (Fig.4b3) TM(OH)$_2$ precursors, no significant agglomeration is observed, while NMC811 from biggest TM(OH)$_2$ precursor of 6.7 µm still displays significantly assembled secondary structures (Fig.4c3), similar to conventional polycrystal NMC811, although the primary particle size is now increased. Comparison of NMC811 developed from 3.6 µm and 4.6 µm TM(OH)$_2$ precursors reveal that NMC811 crystal size distribution is more homogeneous when 4.6 µm precursor is used (Fig.s15). For individual NMC811 single crystals derived from 3.6 µm TM(OH)$_2$ precursors, some of the single crystalline NMC811 is as large as 2.5 µm, although most of as-prepared crystals are at ca.1 µm (Fig.4a3). NMC811 single crystals synthesized from 4.6 µm precursor shows a more uniform and narrow distribution of crystal size (Fig.s15b).

To understand why larger precursors lead to more agglomeration of single crystals, TMO decomposed from 4.6 and 6.7 µm TM(OH)$_2$ precursors are mixed with LiOH separately. The two LiOH/TMO mixtures are heated at different temperatures and compared side-by-side to probe the microstructural evolution of single crystals during chemical lithiation process. At 500°C, the mixture still looks like their TMO precursors because LiOH just starts to melt at 500°C (Fig.s16a1-b1). At 800°C NMC811 begins to form (XRD in Fig.s17) in both cases but with different morphologies observed. Single crystals are more loosely packed in NMC811 prepared from smaller precursor of 4.6 µm (Fig.s16a2), while NMC811 from 6.7 µm precursor is still agglomerated (Fig.s16b2). Because the size of primary particles is the same in all precursors used in this work, the smaller TMO (4.6 µm) has less total number of primary particles compared to 6.7 µm TMO (Fig.4). It will be easier for smaller clusters (4.6 µm TMO) to fully break down to individual ones during crystal growth, aided by redistributed Ni along TMO phase boundaries as discussed earlier. During synthesis, the inconsistent lattice volume changes from rock salt and spinel phases (in TMO) to α-NaFeO$_2$ structure (NMC811) also helps crystal separation [18]. It is more challenging to separate all crystals from the bigger TMO agglomerates (6.7 µm) during reaction with LiOH. This explains why there is still some individual single crystals found in NMC811 prepared from bigger TMO (Fig. 4c3 and Fig.s16b) but most of NMC811 are still "clusters".

The electrochemical properties of NMC811 single crystals from different TM(OH)$_2$ precursors are compared in Fig.s18. Again, the dimensions of all individual single crystals are the same (Fig.4a3-c3), regardless of their precursor sizes. It is the degree of agglomeration that is different among these single crystals from different precursors (Fig.4). NMC811 from 6.7 µm precursor displays the lowest discharge capacity of 185 mAh/g due to the significant agglomeration of micron sized crystals (Fig.s18). NMC811 single crystals derived from 4.6 µm precursors delivers usable capacities at ca.197 mAh/g (Fig.5a), similar to the capacity of NMC811 developed from 3.6 µm precursor (Fig.s18), although the former (from 4.6 µm precursors) displays a more homogeneous crystal size distribution.

This convenient synthesis approach is further utilized to scale up single crystal NMC811 to 200g/batch [19]. Realistic 2 Ah pouch cells employing NMC811 single crystals (from scaleup synthesis using 4.6 µm precursor) as the cathode and graphite anode are prepared to cross validate the materials performances at industry-relevant conditions. The details of large-scale coating of moisture sensitive NMC811 can be found from our recently published work [15]. Double side coated NMC811 cathode has a high mass loading of 18 mg/cm$^2$ corresponding to an areal capacity of 3.5 mAh/cm$^2$ on each side of the current collector. The porosity of the cathode is also controlled at ca. 35%. The initial coulombic efficiency during the first charge-discharge for pouch cells is high at ca. 89.2% (Fig. s19a) without any pre-lithiation of graphite anode. The irreversible loss of Li$^+$ ions is mainly assigned to the formation of solid electrolyte interface layer on graphite anode during formation. After three formation cycles at C/10 (200 mA), the current density of the pouch cell is increased to C/3 (660 mA for both charge and discharge of the pouch cell). A total capacity of 2.03 Ah is delivered from the pouch cell at C/3 (Fig.s19b). An extremely stable cycling has been demonstrated from this Li-ion pouch cell with 69.1% capacity retention over 1000 cycles (Fig. 5b). The extensively cycled single crystal NMC811 harvested from the pouch cell does not indicate obvious cracking nor gliding (Fig. 5c-5d). The bulk phase of cycled single crystals (Fig. 5e and Fig.s20) still maintains as an integrated layered structure after 1000 cycles. On the surface, a very thin (ca.2 nm) and uniform cathode-electrolyte interface (CEI) layer (Fig.5f & Fig.s20) is found, suggesting that a stable CEI further contributes to the observed 1000 stable cycles. Elemental distribution (Fig.5g&5h) also keeps uniform in the extensively cycled crystals. STEM-EELS line scanning (Fig.5i) from bulk region shows the identical peaks indicating homogenous distribution of the electronic structure inside NMC811 single crystal after 1000 cycling. A second pouch cell (Fig.s21) with the same crystal cathode and graphite anode is tested at elevated cutoff voltage of 4.3 V *vs.* graphite (equals to 4.4 V vs. Li/Li$^+$) to further test the stability of single crystal NMC811. After extensive cycling, the single crystals cycled between the expanded electrochemical window are still intact (Fig.s22). No cracking is found from those single crystals charged to high voltages with a trace number of gliding lines identifiable in certain crystal surfaces (Fig.s22d). This observation is consistent with our earlier prediction on the crystal size of 3.5 µm for Ni-rich NMC single crystals, below which cracks will become stable inside the single crystals [2].

**Conclusions**

A scalable solid-state approach for synthesis of single crystal NMC811 is discovered. Decomposition of TM(OH)$_2$ leads to formation of TMO consisting of submicron primary particles benefiting the subsequent growth of single crystal NMC811 after reaction with LiOH. Both rock salt and spinel phases are found in decomposed TMO accompanied by Ni redistribution in TMO that helps deglomeration of single crystal clusters. Equally important is the particle size control of TM(OH)$_2$ precursor which impacts the degree of the agglomeration among single crystals. The benefits from Ni-relocation in TMO and different phase transitions are amplified when relatively smaller TM(OH)$_2$ (and therefore smaller TMO) precursors are employed. NMC811 single crystals derived from 4.6µm precursors demonstrate the balanced crystal size homogeneity (average particle size: 1-2µm) and segregation. When implemented in a 2 Ah Li-ion pouch cells, single crystal NMC811 has displayed an extraordinary cycling stability of 69.1% capacity retention over 1000 cycles. There is no visible cracking identified on the extensively cycled crystals, indicating the structural stability of as-prepared single crystals. The direct synthesis of single crystal NMC811 provides a fundamentally new direction for large-scale production of single crystal cathode materials for advanced lithium-based batteries, inspiring more revolutionary and cost-effective solutions to address the challenges of Ni-rich cathode materials manufacturing.

**References and Notes:**

1 Harlow, J. E. *et al.* A Wide Range of Testing Results on an Excellent Lithium-Ion Cell Chemistry to be used as Benchmarks for New Battery Technologies. *J. Electrochem. Soc.* **166**, A3031-A3044, doi:10.1149/2.0981913jes (2019).
2 Bi, Y. *et al.* Reversible planar gliding and microcracking in a single-crystalline Ni-rich cathode. *Sci* **370**, 1313-1317 (2020).
3 Jung, R. *et al.* Effect of Ambient Storage on the Degradation of Ni-Rich Positive Electrode Materials (NMC811) for Li-Ion Batteries. *J. Electrochem. Soc.* **165**, A132-A141, doi:10.1149/2.0401802jes (2018).
4 Kim, Y., Park, H., Warner, J. H. & Manthiram, A. Unraveling the Intricacies of Residual Lithium in High-Ni Cathodes for Lithium-Ion Batteries. *ACS Energy Lett.* **6**, 941-948, doi:10.1021/acsenergylett.1c00086 (2021).
5 Hu, J. *et al.* Mesoscale-architecture-based crack evolution dictating cycling stability of advanced lithium ion batteries. *Nano Energy* **79**, doi:10.1016/j.nanoen.2020.105420 (2021).
6 Kim, U.-H. *et al.* Heuristic solution for achieving long-term cycle stability for Ni-rich layered cathodes at full depth of discharge. *Nat. Energy* **5**, 860-869, doi:10.1038/s41560-020-00693-6 (2020).
7 Jung, R., Metzger, M., Maglia, F., Stinner, C. & Gasteiger, H. A. Oxygen Release and Its Effect on the Cycling Stability of LiNi$_x$Mn$_y$Co$_z$O$_2$ (NMC) Cathode Materials for Li-Ion Batteries. *J. Electrochem. Soc.* **164**, A1361-A1377, doi:10.1149/2.0021707jes (2017).

8      Li, T. *et al.* Degradation Mechanisms and Mitigation Strategies of Nickel-Rich NMC-Based Lithium-Ion Batteries. *Electrochem. Energy Rev.* **3**, 43-80, doi:10.1007/s41918-019-00053-3 (2019).
9      Yin, S. *et al.* Fundamental and solutions of microcrack in Ni-rich layered oxide cathode materials of lithium-ion batteries. *Nano Energy* **83**, doi:10.1016/j.nanoen.2021.105854 (2021).
10     Yan, P. F. *et al.* Tailoring grain boundary structures and chemistry of Ni-rich layered cathodes for enhanced cycle stability of lithium-ion batteries. *Nat. Energy* **3**, 600-605, doi:10.1038/s41560-018-0191-3 (2018).
11     Liu, Y., Harlow, J. & Dahn, J. Microstructural Observations of "Single Crystal" Positive Electrode Materials Before and After Long Term Cycling by Cross-section Scanning Electron Microscopy. *J. Electrochem. Soc.* **167**, doi:10.1149/1945-7111/ab6288 (2020).
12     Zhang, Y., Liu, J. & Cheng, F. Concentration-Gradient $LiNi_{0.85}Co_{0.12}Al_{0.03}O_2$ Cathode Assembled with Primary Particles for Rechargeable Lithium-Ion Batteries. *Energy Fuels* **35**, 13474-13482, doi:10.1021/acs.energyfuels.1c02115 (2021).
13     Zhu, J. & Chen, G. Y. Single-crystal based studies for correlating the properties and high-voltage performance of $Li[Ni_xMn_yCo_{1-x-y}]O_2$ cathodes. *Journal of Materials Chemistry A* **7**, 5463-5474, doi:10.1039/c8ta10329a (2019).
14     Qian, G. *et al.* Single-crystal nickel-rich layered-oxide battery cathode materials: synthesis, electrochemistry, and intra-granular fracture. *Energy Storage Materials* **27**, 140-149, doi:10.1016/j.ensm.2020.01.027 (2020).
15     Bi, Y., Li, Q., Yi, R. & Xiao, J. To pave the way for large-scale electrode processing of moisture-sensitive Ni-rich cathodes. *J. Electrochem. Soc.* **169**, 020521, doi:10.1149/1945-7111/ac4e5d (2022).
16     Lee, M. H., Kang, Y. J., Myung, S. T. & Sun, Y. K. Synthetic optimization of $Li[Ni_{1/3}Co_{1/3}Mn_{1/3}]O_2$ via co-precipitation. *Electrochim. Acta* **50**, 939-948, doi:10.1016/j.electacta.2004.07.038 (2004).
17     Ou, X. *et al.* Enabling high energy lithium metal batteries via single-crystal Ni-rich cathode material co-doping strategy. *Nat. Commun.* **13**, doi:10.1038/s41467-022-30020-4 (2022).
18     Wang, S. *et al.* Kinetic Control of Long‐Range Cationic Ordering in the Synthesis of Layered Ni‐Rich Oxides. *Adv. Funct. Mater.* **31**, 2009949, doi:10.1002/adfm.202009949 (2021).
19     Xiao, J. Scaling up of high performance single crystalline Nicel-rich cathode materials with advanced litium salts, presented at the U.S. Department of Energy Vehicle Technologies Office Annual Merit Review, BAT552, 22 June 2022.

**Acknowledgments:** The author thanks Dr. L. Jiang, Dr. Z. Liu at Thermal Fisher Scientific for their help on TEM characterizations. This research used beamline 28-ID-2 (XPD) of the National Synchrotron Light Source II, U.S. DOE Office of Science User Facilities, operated for the DOE Office of Science by Brookhaven National Laboratory under contract no. DE-SC0012704. **Funding:** This work was supported by the Assistant Secretary for Energy Efficiency and Renewable Energy (EERE), Advanced Manufacturing Office and jointly supported by Vehicle Technology Office of the US Department of Energy (DOE) under Award Number DE-LC-000L080. E.H. and S.T. are supported by DOE/EERE/VTO through the Advanced Battery Materials Research program under contract no. DE-SC0012704. **Author contributions:** J.X. and Y.B. conceptualized the idea and proposed the research. Y.B. synthesized materials with the help from R.Y. Y.X., Z.P., C.W. J. H did characterizations and sample preparation for this work. Y.B., D.L. and Q.L. assembled pouch cells. S.T. and E.H. helped on XRD and structural analysis. J.L., R.O., L.L., X.H., S.V. and J.R. provided lithium salts for this work and helped analyze the materials and results through collaboration project. J.X. drafted the manuscript with input from Y.B., Y.X and R.Y. All authors helped editing and modifying the manuscript. **Competing interests:** The authors declare no competing interests. **Data and materials availability:** All data are available in the main text or supplementary materials.

**Supplementary Materials:**

Materials and Methods

Figures s1-s22

**Fig. 1.** Schematic of solid-state synthesis approach for single crystal Ni rich cathode.

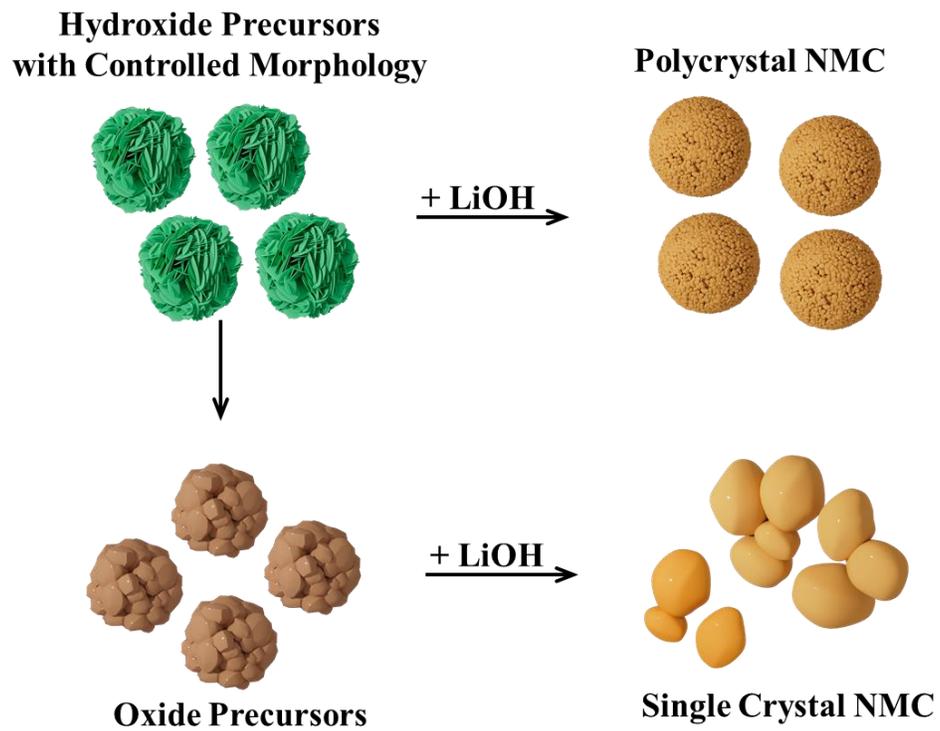

**Fig. 2.** Morphology, phase analysis and element mapping of TM(OH)$_2$, TMO precursors and as-formed single crystals. a, Cross sectional SEM image of TM(OH)$_2$ precursor indicates voids and pores in its secondary structures. b, Cross sectional SEM image of TMO precursor reveals increased size of primary particles to submicron. c, High-angle annular dark-field scanning transmission electron microscopy (HAADF) image identifies two mixed phases, rock salt and spinel, in TMO. Red inset is Fast Fourier transform (FFT) image of rock salt region. Orange inset is FFT image of spinel phase region. d, Cross sectional SEM image of single crystal NMC811 indicates dense structure without any grain boundary across the entire crystal. e, HAADF image of single crystalline NMC811 and EDS mapping shows uniform distribution of Ni, Co and Mn. f, STEM image of single crystal NMC811. g, Integrated differential phase contrast (iDPC) image of single crystal NMC811.

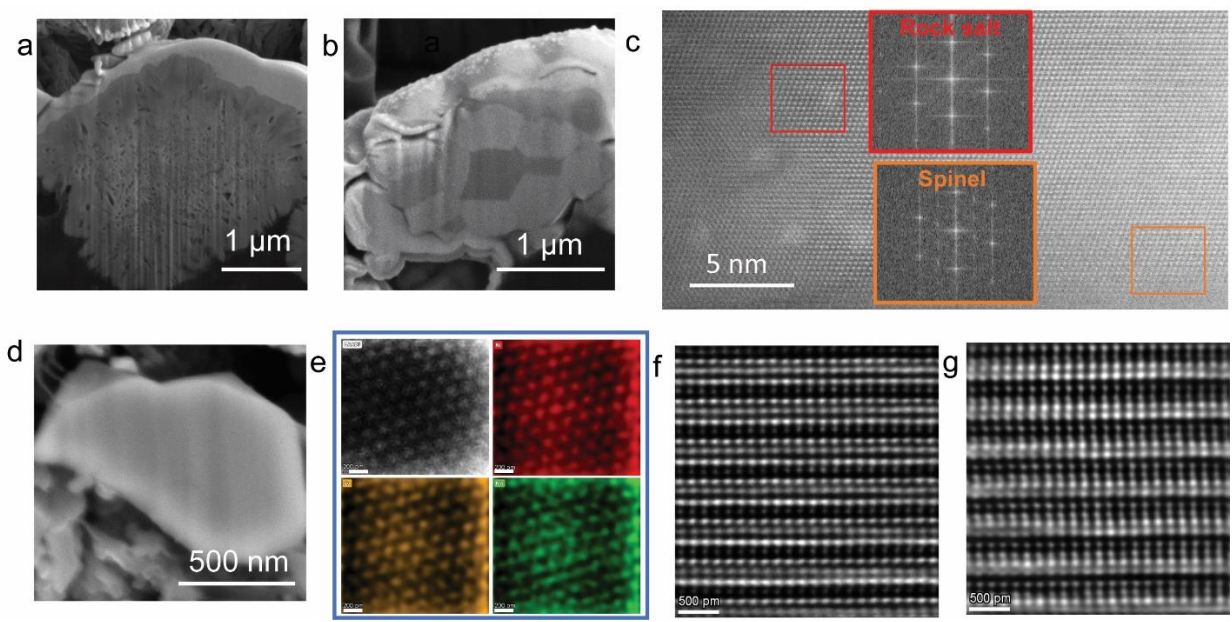

**Fig. 3.** Structure and phase evolution of TM(OH)$_2$ decomposed at different temperatures to form TMO. SEM images of TMO derived from (a1) 500°C, (b1) 600°C (c1) 700°C (d1) 800°C and (e1) 900°C show that the primary particles gradually transform from nanoplates to submicron sized particles in TMO with rising temperatures. (a2, b2, c2, d2, e2) are the corresponding cross-sectional STEM images of TMO precursors in (a1, b1, c1, d1 and e1). Selected area electron diffraction (SAED) image of TMO precursors derived from 500°C-900°C are compared in (a3, b3, c3, d3, e3) which indicates mixed phases of rock salt and spinel in TMO at all different temperatures. (a4, b4, c4, d4, e4) are the STEM-EDS mapping of TMO precursors derived from different temperatures. Ni segregation becomes prominent in TMO as temperature increases as a result of the co-existence of both rock salt and spinel phases in TMO. a1-a4, b1-b4, c1-c4, d1-d4 and e1-e4 correspond to TMO calcinated at 500, 600, 700, 800, and 900 °C, respectively.

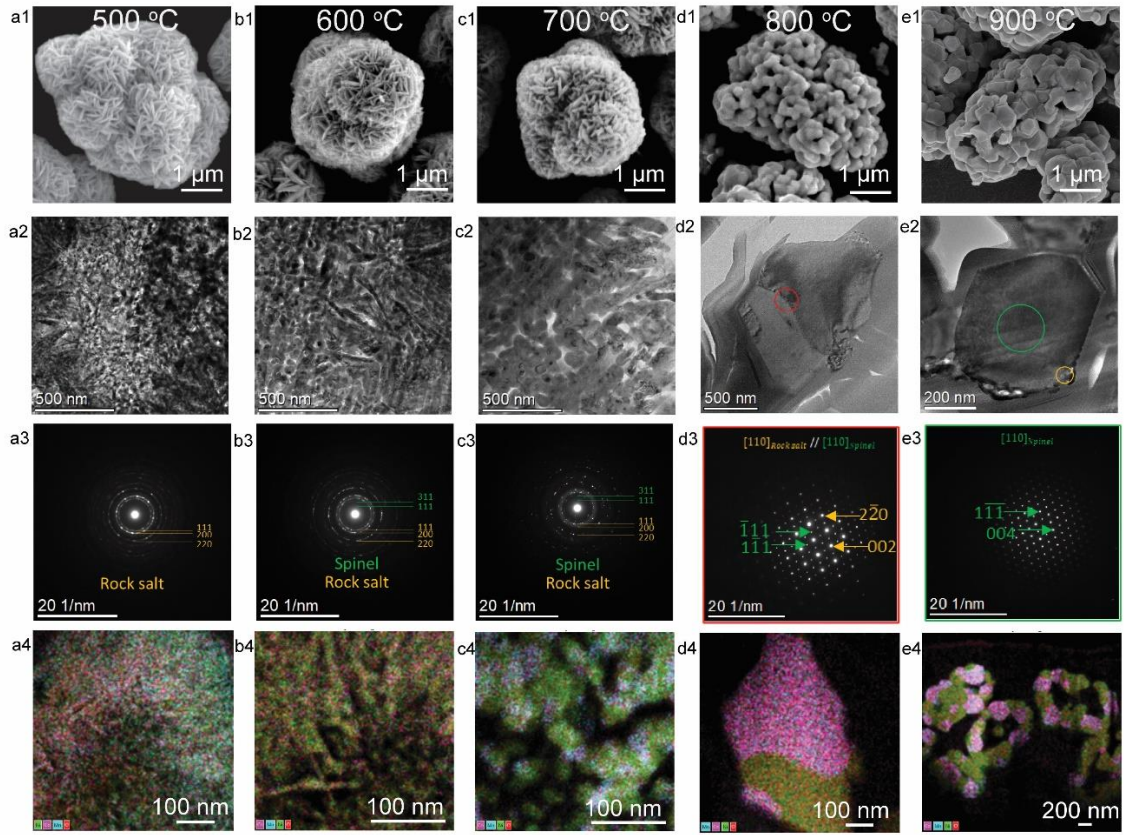

**Fig. 4**. SEM images to examine the impacts of TM(OH)$_2$ precursor size on the degree of agglomeration of as-synthesized single crystal NMC811. TM(OH)$_2$ precursors with different particle sizes of (a1) 3.6 µm, (b1) 4.6 µm, (c1) 6.7 µm are decomposed to corresponding TMO intermediates in (a2, b2 and c2). After reaction with LiOH, different morphologies are uncovered (a3, b3 and c3) for NMC811 single crystals derived from various TMO precursors in (a2-c2).

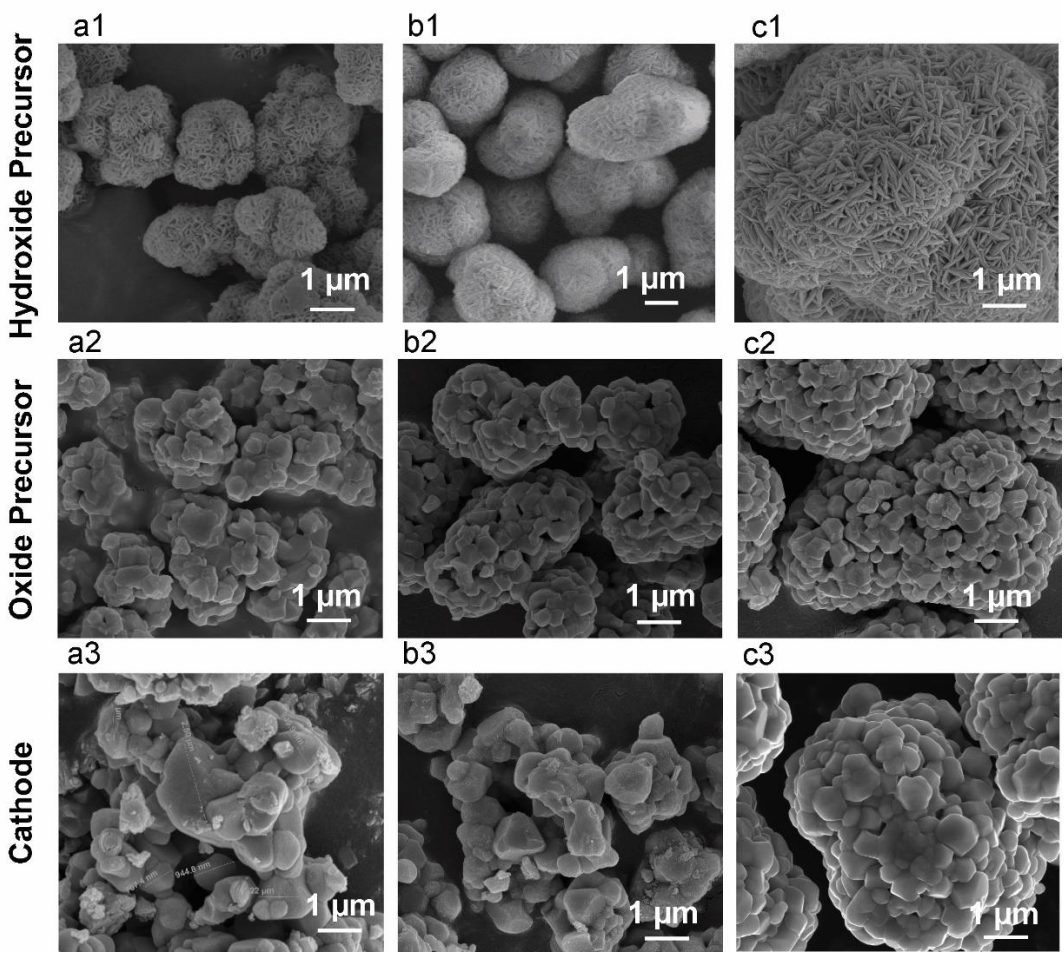

**Fig. 5.** Validation of as-synthesized single crystal NMC811 in a 2Ah Li-ion pouch cell. a, The initial charge-discharge curve of the single crystal NMC811 at C/10 rate in half cell (2.7-4.4V *vs.* Li$^+$/Li). b, The prototype 2 Ah pouch cell consisting of as-prepared single crystal NMC811 cathode and graphite anode demonstrates 1000 stable cycling with 69.1% capacity retention. Electrochemical window: 2.6-4.2V vs. graphite. Rate: C/3 (660mA). Cathode areal capacity of 3.5 mAh/cm$^2$ (18 mg/cm$^2$). Three formation cycles were conducted at C/10 (200mA). (c) SEM image, (d) cross-sectional STEM images of single crystal NMC811 harvested from the pouch cell after 1000 cycles still display integrated crystal structures without gliding or cracking. STEM-HAADF image on (e) bulk and (f) surface area of 1000 cycled single crystal. (g) (h) STEM-EDS image of cycled single crystal. (i) Electron energy loss spectroscopy (EELS) collected along the arrow in (g).

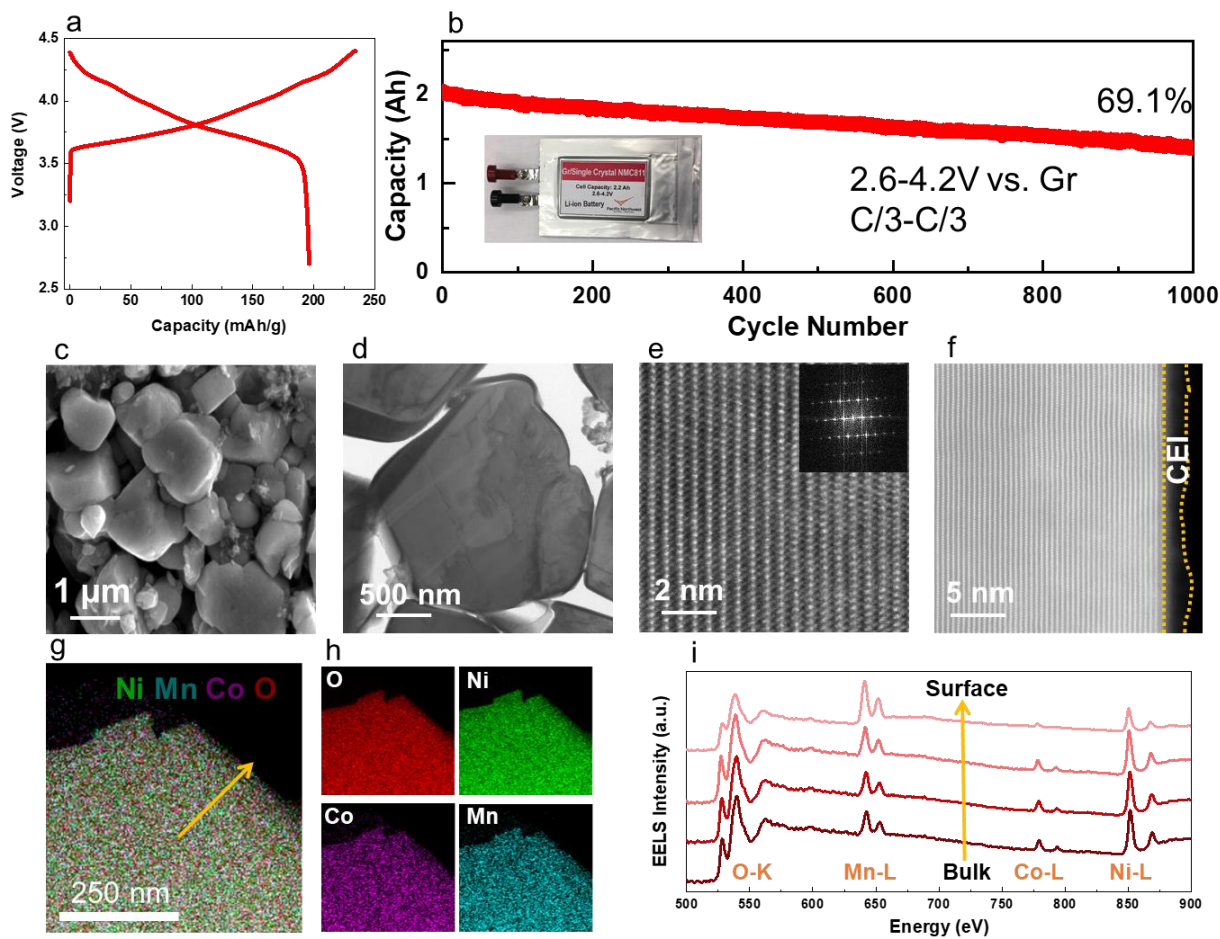